\documentclass[a4paper,10pt,twoside]{cpc-hepnp}
\usepackage{CJK,upgreek,fancyhdr}
\usepackage{multicol}
\usepackage{graphicx}
\usepackage{booktabs}
\usepackage{amssymb,bm,mathrsfs,bbm,amscd}
\usepackage[tbtags]{amsmath}
\usepackage{lastpage}
\usepackage{longtable}
\usepackage{changes}

\begin{document}
\begin{CJK*}{GB}{gbsn}

\fancyhead[c]{\small Chinese Physics C~~~Vol. xx, No. x (201x) xxxxxx}
\fancyfoot[C]{\small 010201-\thepage}


\title{$\boldsymbol{\mathcal{\alpha}}$ decay properties of $\boldsymbol{^{296}}$Og within the two-potential approach
\thanks{Supported by the National Natural Science Foundation of China (Grants No. 11205083 and No. 11505100), the Construct Program of the Key Discipline in Hunan Province, the Research Foundation of Education Bureau of Hunan Province, China (Grant No. 15A159), the Natural Science Foundation of Hunan Province, China (Grants No. 2015JJ3103 and No. 2015JJ2121), the Innovation Group of Nuclear and Particle Physics in USC, the Shandong Province Natural Science Foundation, China (Grant No. ZR2015AQ007), Hunan Provincial Innovation Foundation For Postgraduate (Grant No. CX2017B536).}}

\author{%
   Jun-Gang Deng (µË¾ü¸Õ)$^{1}$
\quad Jie-Cheng Zhao (ÕԽ׳É)$^{1}$
\quad Jiu-Long Chen (³Â¾ÁÁú)$^{1}$\\
\quad Xi-Jun Wu (Îâϲ¾ü)$^{2;1)}$\email{{wuxijun1980@yahoo.cn}}
\quad Xiao-Hua Li (ÀîС»ª)$^{1,3,4;2)}$\email{lixiaohuaphysics@126.com}%
}
\maketitle

\address{%
$^1$ School of Nuclear Science and Technology, University of South China, Hengyang 421001, China\\
$^2$ School of Math and Physics, University of South China, Hengyang 421001, China\\
$^3$ Cooperative Innovation Center for Nuclear Fuel Cycle Technology $\&$ Equipment, University of South China, Hengyang 421001, China\\
$^4$ Key Laboratory of Low Dimensional Quantum Structures and Quantum Control, Hunan Normal University, Changsha 410081, China\\
}

\begin{abstract}
The present work is a continuations of our previous paper [J.-G. Deng, et al., Chin. Phys. C, {\bf41}: 124109 (2017)]. In present work, the $\mathcal{\alpha}$ decay half-life of unknown nucleus $^{296}$Og is predicted within the two-potential approach and the hindrance factors of all 20 even-even nuclei in the same region with $^{296}$Og, i.e. proton number $82<Z<126$ and neutron number $152<N<184$, from $^{250}$Cm to $^{294}$Og are extracted. The prediction is 1.09 ms within a factor of 5.12. In addition, based on the latest experimental data, a new set of parameters of $\mathcal{\alpha}$ decay hindrance factors for the even-even nuclei in this region considering the shell effect and proton-neutron interaction are obtained.
\end{abstract}



\footnotetext[0]{\hspace*{-3mm}\raisebox{0.3ex}{$\scriptstyle\copyright$}2016
Chinese Physical Society and the Institute of High Energy Physics
of the Chinese Academy of Sciences and the Institute
of Modern Physics of the Chinese Academy of Sciences and IOP Publishing Ltd}%

\begin{multicols}{2}

\section{Introduction}

The synthesis and identification of superheavy nuclei have been a hot topic in nuclear physics since the prediction of the existence of a superheavy island in the 1960s \cite{SOBICZEWSKI1966500,NILSSON19691,Mosel1969,Meldner_1967,PhysRevC.94.054621}. During more than ten years, the superheavy nuclei, from $Z$ = 113 to $Z$ = 118, have been synthesized by hot-fusion reactions between $^{48}$Ca beams and radioactive actinide targets \cite{PhysRevC.76.011601,PhysRevC.74.044602,0954-3899-34-4-R01,PhysRevLett.104.142502,PhysRevLett.105.182701,Hofmann2012}. Presently, the synthesis of $^{296}$Og is excepted to be via the reaction $^{251}$Cf($^{48}$Ca, 3n)$^{296}$Og at the Flerov Laboratory of Nuclear Reactions (FLNR) in Dubna, Russia \cite{doi:10.1063/PT.3.2880,PhysRevC.94.051302}. If the experiment succeeds, nucleus $^{296}$Og will be the nucleus observed with the largest number of protons and neutrons, which is the nucleus closest to predicted $N$=184 shell closure \cite{SOBICZEWSKI1966500,Mosel1969}. Spontaneous fission and $\mathcal{\alpha}$ decay are the two main decay modes of superheavy nuclei. For superheavy nuclei around Rf, $\mathcal{\alpha}$ decay is a weaker candidate compared to spontaneous fission \cite{PhysRevC.94.054621}. For the most of recently synthesized proton-rich ones, $\mathcal{\alpha}$ decay is the dominant decay mode \cite{PhysRevC.94.054621}. Recently, Bao et al. \cite{PhysRevC.95.034323} also predicted that $\mathcal{\alpha}$ decay is the main decay mode for $^{296}$Og.

An accurate prediction of $\mathcal{\alpha}$ decay half-life will be used as a reference on experiment synthetizing nucleus $^{296}$Og. Recently, Sobiczewski \cite{PhysRevC.94.051302} predicted $\mathcal{\alpha}$ decay half-life $T_{1/2}$ of $^{296}$Og by adopting a 3-parameter phenomenological formula of $T_{1/2}$ \cite{2005AcPPB..36.3095P}, while the $\mathcal{\alpha}$ decay energy $Q_{\alpha}$ are obtained from nine different mass models as belows: M$\ddot{\rm{o}}$ller et al. (FRDM) \cite{MOLLER1995185}, Duflo and Zuker (DZ) \cite{PhysRevC.52.R23}, Nayak and Satpathy (INM) \cite{Nayak2012616}, Wang and Liu (WS3+) \cite{PhysRevC.84.051303}, Wang et al. (WS4+) \cite{PhysRevC.93.014302,Wang2014215}, Muntian et al. (HN) \cite{2001AcPPB..32..691M, Sobiczewski2007292}, Kuzmina et al. (TCSM) \cite{PhysRevC.85.014319}, Goriely et al. (HFB31) \cite{PhysRevC.93.034337}, and Liran et al. (SE) \cite{PhysRevC.62.047301}. For the purpose of obtaining a more accurate $T_{1/2}$ of $^{296}$Og, how to select a more precise $Q_{\alpha}$ is one of the heart of the matters. Sobiczewski \cite{PhysRevC.94.051302} found that the deviation between measured $\mathcal{\alpha}$ decay half-life and calculation, adopting $Q_{\alpha}$ from WS3+ \cite{PhysRevC.84.051303}, was minimal, by analyzing calculations of nine different mass models. However, the parameters of the phenomenological formula \cite{2005AcPPB..36.3095P} adopted by Sobiczewski, were extracted from NUBASE2003 \cite{AUDI20033}, and AME2003 \cite{WAPSTRA2003129,AUDI2003337} for nuclei with $Z$=84--110, and $N$=128--160. Very recently, Mohr \cite{PhysRevC.95.011302} adopted the systematic behavior of strength parameter for the double-folding potential of $\mathcal{\alpha}$-core to predict the $Q_{\alpha}$ and $T_{1/2}$ of $^{296}$Og. The predicted results are $Q_{\alpha}$=11.655 $\pm$ 0.095 MeV and $T_{1/2}$ = 0.825 ms with an uncertainty factor of 4.

In our previous works \cite{PhysRevC.93.034316,PhysRevC.95.044303,PhysRevC.95.014319,CPC-124109}, we used the two-potential approach(TPA) \cite{PhysRevLett.59.262,PhysRevA.69.042705} to systematically study the $\mathcal{\alpha}$ decay hindrance factors and/or preformation probabilities for even-even, odd-$A$ and doubly-odd nuclei, and we found that the behaviors of the $\mathcal{\alpha}$ decay hindrance factors and/or preformation probabilities of the same kinds nuclei(even-even nuclei, odd-$A$ nuclei and doubly-odd nuclei) in the same region are similar, while the regions are divided by the magic numbers of proton and neutron. In present work, in order to reduce the uncertainty factor of prediction of nucleus $^{296}$Og, we systematically study all of 20 even-even nuclei of 82 $< Z <$ 126 and 152 $<N<$ 184 in the same region with $^{296}$Og from $^{250}$Cm to $^{294}$Og. The $\mathcal{\alpha}$ decay energies and half-lives are taken from the latest evaluated nuclear properties table NUBASE2016 \cite{1674-1137-41-3-030001} and evaluated atomic mass table AME2016 \cite{1674-1137-41-3-030002,1674-1137-41-3-030003} except the $Q_{\alpha}$ of nucleus $^{296}$Og is from WS3+ \cite{PhysRevC.84.051303}.

This article is organized as follows. In next section, the theoretical framework of calculating $\mathcal{\alpha}$ decay half-life is briefly described. The detailed calculations and discussion are presented in Section 3. Finally, a summary is given in Section 4.

\section{Theoretical framework}
\label{section 2}

The TPA \cite{PhysRevLett.59.262, PhysRevA.69.042705} was put forward to investigate the quasi-stationary problems, initially. Recently, it is widely used to deal with $\mathcal{\alpha}$ decay \cite{PhysRevC.94.024338,1674-1137-41-1-014102,PRC-024318,PhysRevC.85.027306, QIAN201182, QIAN20111}. In the framework of the TPA, the $\mathcal{\alpha}$ decay half-life $T_{1/2}$ is calculated by
\begin{equation}
\
T_{1/2}=\frac{{\hbar}ln2}{\Gamma}=\frac{ln2}{\lambda}
,\label{subeq:1}
\end{equation}
where $\hbar$, $\Gamma$ and $\lambda$ denote the Planck constant, decay width and decay constant, respectively. $\lambda$ is depended on the $\mathcal{\alpha}$ particle preformation factor $P_0$, the penetration probability $P$, and the normalized factor $F$. It can be expressed by
\begin{equation}
\
\lambda=\frac{{\hbar}P_0FP}{4{\mu}h}
,\label{subeq:2}
\end{equation}
where $\mu=\frac{{m_d}{m_{\alpha}}}{{m_d}+{m_{\alpha}}}$ is the reduced mass between daughter nucleus and preformed $\mathcal{\alpha}$ particle with the mass of daughter nucleus $m_d$ and $\mathcal{\alpha}$ particle $m_{\alpha}$. $P_0$ is the $\mathcal{\alpha}$ preformation factor, on account of the complicated structure of quantum many-body systems, there are a few of works \cite{QI201677,PhysRevC.77.054318,PhysRevC.92.044302,PhysRevC.93.011306,PhysRevC.95.061306} studying $P_0$ from the viewpoint of microscopic theory. In accordance with the calculations by adopting the density-dependent cluster model (DDCM) \cite{XU2005303}, $P_0$ is 0.43 for even-even nuclei. 

$h$ is the hindrance factor, denoting the deviation between calculation and experimental data of $\mathcal{\alpha}$ decay half-life, which can be expressed as
\begin{equation}
\
h=\frac{T_{1/2}^{\rm{exp}}}{T_{1/2}^{\rm{cal}}}
,\label{3}
\end {equation}
where $T_{1/2}^{\rm{exp}}$ and $T_{1/2}^{\rm{cal}}$ are $\mathcal{\alpha}$ decay half-live of experimental data and calculated value with $P_0=0.43$ \cite{XU2005303}.
Recently, a simple formula, considering the nuclear shell effect and proton-neutron interaction for estimating the $\mathcal{\alpha}$ decay hindrance factors\rq{} variation tendency, was putted forward \cite{PhysRevC.93.034316,PhysRevC.80.057301,GUO2015110} and written as
\begin{eqnarray}
\log_{10}h=&a+b(Z-Z_1)(Z_2-Z)+c(N-N_1)(N_2-N)\nonumber\\
&+ dA+e(Z-Z_1)(N-N_1).\label{4} 
\end{eqnarray}
where $Z$ and $N$ are the proton and neutron number of parent nucleus. $Z_1$ ($N_1$) and $Z_2$ ($N_2$) denote the proton (neutron) magic numbers with $Z_1<Z<Z_2$ and $N_1<N<N_2$. $a$, $b$, $c$, $d$ and $e$ are adjustable parameters.

The barrier penetrate probability $P$ is obtained by the semiclassical Wentzel-Kramers-Brillouin (WKB) approximation and written as
\begin{equation}
\
P=\exp(-2{\int_{r_2}^{r_3} k(r) dr})
,\label{subeq:3}
\end{equation}
where $k(r)=\sqrt{\frac{2\mu}{{\hbar}^2}|Q_{\alpha}-V(r)|}$ represents the wave number of the $\mathcal{\alpha}$ particle. $r$ is mass center distance between $\mathcal{\alpha}$ particle and daughter nucleus. $V(r)$ denotes the entire $\mathcal{\alpha}$-core potential.

The normalized factor $F$, indicating the assault frequency of $\mathcal{\alpha}$ particle, can be approximatively obtained by
\begin{equation}
\
F{\int_{r_1}^{r_2} \frac{1}{2k(r)} dr}=1,
\label{subeq:1}
\end{equation}
where $r_1$, $r_2$ and the foregoing $r_3$ are the classical turning points, they satisfy the conditions $V (r_1) = V (r_2) = V (r_3) =Q_\alpha$.

The entire $\mathcal{\alpha}$-core potential $V(r)$, which is composed with the nuclear potential $V_N(r)$, the Coulomb potential $V_C(r)$, and the centrifugal potential $V_l(r)$, is expressed as
\begin{equation}
\
V(r)=V_N(r)+V_C(r)+V_l(r)
.\label{subeq:1}
\end {equation}
In this work, we choose a type of cosh parametrized form for $V_{N}(r)$, obtained by analyzing experimental data of $\mathcal{\alpha}$ decay \cite{PhysRevLett.65.2975}, which is written as
\begin{equation}
\
V_N(r)=-V_0\frac{1+\text{cosh}(R/a_0)}{\text{cosh}(r/a_0)+\text{cosh}(R/a_0)},
\label{subeq:1}
\end {equation}
where $V_0$ and $a_0$ denote the depth and diffuseness of the nuclear potential. In our past work \cite{PhysRevC.93.034316}, we have obtained a set of parameters considering the isospin effect, which is $a_0=0.5958$ fm and $V_0=192.42+31.059\frac{N_d-Z_d}{A_d}$ MeV. Here $N_d$, $Z_d$ and $A_d$ denote the neutron, proton and mass number of the daughter nucleus, respectively. The nuclear potential sharp radius $R$ is calculated by the nuclear droplet model and proximity energy \cite{0954-3899-26-8-305} with the mass number of parent nucleus $A$, and written as
\begin{equation}
\
R=1.28A^{1/3}-0.76+0.8A^{-1/3}
.\label{subeq:1}
\end {equation}
The Coulomb potential $V_C(r)$ is taken as the potential of a uniformly charged sphere with sharp radius $R$, which is expressed as follows
\begin{equation}
\
V_C(r)=\left\{\begin{array}{ll}

\frac{Z_dZ_{\alpha}e^2}{2R}[3-(\frac{r}{R})^2],&\text{{r}\textless{R}},\\

\frac{Z_dZ_{\alpha}e^2}{r},&\text{{r}\textgreater{R}},

\end{array}\right.
\label{subeq:1}
\end {equation}
where $ Z_{\alpha}=2$ denotes the proton number of $\mathcal{\alpha}$ particle.

As for the centrifugal barrier $V_l(r)$, we adopt the Langer modified form, because $l(l+1){\to}(l+1/2)^2$ is a necessary correction for one-dimensional problems \cite{1995JMP....36.5431M}. It can be expressed as
\begin{equation}
\
V_l(r)=\frac{{\hbar}^2(l+1/2)^2}{2{\mu}r^2}
,\label{subeq:1}
\end {equation}
where $l$ denotes the orbital angular momentum taken away by the $\mathcal{\alpha}$ particle. $l=0$ for the favored $\mathcal{\alpha}$ decays, while $l{\ne}0$ for the unfavored decays. In the case of even-even nuclei $\mathcal{\alpha}$ decay, $l=0$.

\section{Results and discussion}

Recently, Yao et al. \cite{EPJA-122} used fourteen different versions of proximity potentials to calculate $\mathcal{\alpha}$ decay half-life. Their research shows that the results of the generalized proximity potential 1977 (gp77) \cite{PRC-064611} are very in agreement with the experimental data and the gp77 is the most suitable one for calculating the $\mathcal{\alpha}$ decay half-life. For the purpose of obtaining a precise prediction of $\mathcal{\alpha}$ decay half-life for $^{296}$Og, we systematically study all of 20 even-even nuclei in the same region with $^{296}$Og from $^{250}$Cm to $^{294}$Og by adopting the TPA. In addition, for comparisons, we also calculate the $\mathcal{\alpha}$ decay half-lives of these nuclei by adopting gp77 \cite{EPJA-122,PRC-064611}. The calculations include two $\mathcal{\alpha}$ decay chains: the known chain $^{294}$Og $\to{}^{290}$Lv $\to{}^{286}$Fl $\to{}^{282}$Cn, and the chain $^{296}$Og $\to{}^{292}$Lv $\to{}^{288}$Fl $\to{}^{284}$Cn, where decay modes of the later three nuclei are known. $^{282}$Cn and $^{284}$Cn decay only by spontaneous fission and end chains.

Firstly, we calculate $\alpha$ decay half-lives taking $P_0=0.43$ \cite{XU2005303} within TPA for all of 20 even-even nuclei in the same region with nucleus $^{296}$Og, and obtain the corresponding hindrance factors $h$ by Eq. (\ref{3}). Furtherly, based on the obtained $h$ and Eq. (\ref{4}), we fit and extract the corresponding parameters $a$, $b$, $c$, $d$ and $e$, and list in Table \ref{table1}, where in this region, $82<Z\leq{126}$ and $152<N\leq{184}$, $Z_1=82$, $Z_2=126$, $N_1=152$, $N_2=184$. In our previous work \cite{PhysRevC.93.034316}, we have obtained a set of parameters for this region. In this wok, based on the latest experimental data of NUBASE2016 \cite{1674-1137-41-3-030001} and AME2016 \cite{1674-1137-41-3-030002,1674-1137-41-3-030003}, we renewedly extracted a set of parameters contraposing this region. The standard deviation $\sigma_{\rm{pre}}=\sqrt{\sum ({\log_{10}T^{\rm{pre}}_{1/2}-\log_{10}T^{\rm{exp}}_{1/2}})^2/n}$ denotes deviations of $\mathcal{\alpha}$ decay half-live between predictions considering the hindrance factor correction and experimental data for these 20 even-even nuclei. The values of $\sigma_{\rm{pre}}$ drops from 0.32 by adopting the parameters of past work \cite{PhysRevC.93.034316} to 0.26 by using the new ones. It indicates that the predictions adopting new parameters improve by $\frac {0.32-0.26}{0.32}=18.75\%$, where $T^{\rm{pre}}_{1/2}=h^**T_{1/2}^{\rm{cal}}$ with $h^*$ obtained by Eq. (\ref{4}) and $T^{\rm{cal}}_{1/2}$ taking $P_0=0.43$ \cite{XU2005303}.

$\mathcal{\alpha}$ decay energy is an important input for calculating $\mathcal{\alpha}$ decay half-life. Sobiczewski \cite{PhysRevC.94.051302} found that the calculation taking $\mathcal{\alpha}$ decay energy from WS3+ \cite{PhysRevC.84.051303} can best reproduce experimental $\mathcal{\alpha}$ decay half-life. In present work, we select $\mathcal{\alpha}$ decay energy from WS3+ \cite{PhysRevC.84.051303} to calculate half-life of $^{296}$Og. In order to verify the accuracy of the WS3+ \cite{PhysRevC.84.051303} and obtain the errors caused by uncertainty of the $\mathcal{\alpha}$ decay energy, we use $\mathcal{\alpha}$ decay energy from WS3+ \cite{PhysRevC.84.051303} to calculate the $\mathcal{\alpha}$ decay half-life within TPA for 20 even-even nuclei in the same region with $^{296}$Og. The detailed calculations are given in Table \ref{table2}. In this table, the first third columns denote $\mathcal{\alpha}$ decay, experimental data of $\mathcal{\alpha}$ decay energy and half-life. The fourth and fifth columns are $\mathcal{\alpha}$ decay energy from WS3+ \cite{PhysRevC.84.051303} denoted as $Q_{\alpha}^{\rm{WS3+}}$ and calculated half-life taking $P_0=0.43$ \cite{XU2005303} and $Q_{\alpha}^{\rm{WS3+}}$ within TPA denoted as $T_{1/2}^{\rm{WS3+}}$. The sixth and seventh columns are the $\mathcal{\alpha}$ decay half-life calculated by adopting TPA taking $P_0=0.43$ \cite{XU2005303} and experimental decay energy denoted as $T_{1/2}^{\rm{cal}}$, and extracted hindrance factor by Eq. (\ref{3}) denoted as $h$. The eighth column denotes the hindrance factor $h^*$, considering the shell effect and proton-neutron interaction, which is calculated within Eq. (\ref{4}) while the parameters listed in Table \ref{table1}. The ninth one is theoretical prediction of experimental half-life by $T^{\rm{pre}}_{1/2}=h^**T_{1/2}^{\rm{cal}}$. The last one is the calculation by adopting the gp77 \cite{EPJA-122,PRC-064611} denoted as $T_{1/2}^{\rm{gp77}}$.

From Table \ref{table2}, we can clearly see that the $Q_{\alpha}^{\rm{WS3+}}$ and $T^{\rm{WS3+}}_{1/2}$ can reproduce $Q_{\alpha}$ and $T^{\rm{exp}}_{1/2}$ well, respectively. It confirms our confidence to calculate the $\mathcal{\alpha}$ decay energy of $^{296}$Og using WS3+ . In addition, we can see that the $T^{\rm{pre}}_{1/2}$ and $T^{\rm{cal}}_{1/2}$ are better to reproduce the experimental data $T^{\rm{exp}}_{1/2}$ than $T^{\rm{gp77}}_{1/2}$ especially for cases of nuclei $^{250}$Cm, $^{252}$Cf, $^{254}$Cf, $^{256}$Cf and $^{256}$Fm, and $T^{\rm{pre}}_{1/2}$ is the most accurate. Furthermore, we calculate the standard deviation $\sigma_{\rm{WS3+}}=\sqrt{\sum ({\log_{10}T^{\rm{WS3+}}_{1/2}-\log_{10}T^{\rm{exp}}_{1/2}})^2/n}=0.68$, $\sigma_{\rm{cal}}=\sqrt{\sum ({\log_{10}T^{\rm{cal}}_{1/2}-\log_{10}T^{\rm{exp}}_{1/2}})^2/n}=0.46$, $\sigma_{\rm{gp77}}=\sqrt{\sum ({\log_{10}T^{\rm{gp77}}_{1/2}-\log_{10}T^{\rm{exp}}_{1/2}})^2/n}=0.69$ between $T^{\rm{WS3+}}_{1/2}$, $T^{\rm{cal}}_{1/2}$, $T^{\rm{gp77}}_{1/2}$ and $T^{\rm{exp}}_{1/2}$, respectively. The value of $\sigma_{\rm{cal}}$, $\sigma_{\rm{gp77}}$ are larger than $\sigma_{\rm{pre}}$. Therefore $T_{1/2}^{\rm{pre}}$ improve $\frac{0.46-0.26}{0.46}=43.48\%$, $\frac{0.69-0.26}{0.69}=62.32\%$ compared to $T_{1/2}^{\rm{cal}}$ and $T_{1/2}^{\rm{gp77}}$, respectively.

The experimental data and predicted results are plotted as logarithmic forms in Fig. \ref{figure1}. In this figure, the blue circle and red star denote the experimental half-lives $T_{1/2}^{\rm{exp}}$, and predicted results $T_{1/2}^{\rm{pre}}$, respectively. From Fig. \ref{figure1}, we can see that the predicted half-lives are almost equal to the corresponding experimental data. In order to intuitively survey the deviations of those, we plot the logarithms differences between predictions and experimental data in Fig. \ref{figure2}. From this figure, we can clearly see that the values of $\log_{10}{T^{\rm{pre}}_{1/2}}-\log_{10}{T^{\rm{exp}}_{1/2}}$ are around 0, indicating our predictions can well reproduce the experimental data. Therefore, extending our study to predict the $\mathcal{\alpha}$ decay half-life and hindrance factor of nucleus $^{296}$Og are believable. The standard deviation caused by $Q_{\alpha}^{\rm{WS3+}}$ and $h^*$ are $\sigma_{\rm{WS3+}}=0.68$ and $\sigma_{\rm{pre}}$=0.26. We assume that the impact of above errors are equally, thus, the predicted half-life of $^{296}$Og is 1.09 ms within a factor of $\sqrt{({10^{0.68}})^2+({10^{0.26}})^2}$=5.12.

\begin{center}
\tabcaption{ \label{table1} The parameters of $\mathcal{\alpha}$ decay hindrance factor for even-even nuclei from $82<Z\leq126$ and $152<N\leq184$.}
\footnotesize
\begin{tabular*}{80mm}{c@{\extracolsep{\fill}}ccccc}
\toprule $a$&$b$&$c$&$d$&$e$\\
\hline
-24.4069&0.0017&-0.0010&0.0935&-0.0036\\
\bottomrule
\end{tabular*}
\vspace{0mm}
\end{center}
\vspace{0mm}

\end{multicols}

\begin{center}
\tabcaption{The calculated results of $T_{1/2}$, $Q_{\alpha}$, $h$ and $h^*$ with the $Q_{\alpha}$ and $T^{\text{exp}}_{1/2}$ from NUBASE2016 \cite{1674-1137-41-3-030001} and AME2016 \cite{1674-1137-41-3-030002, 1674-1137-41-3-030003} as well as $Q_{\alpha}^{\text{WS3+}}$ from WS3+ \cite{PhysRevC.84.051303} for the 20 even-even nuclei in the same region with nucleus $^{296}$Og. All the $\mathcal{\alpha}$ decay energy and half-lives are in the unit of \lq{}MeV\rq{} and \lq{}s\rq{}, respectively.}
\footnotesize
\label{table2}
\begin{longtable}{cccccccccc}
\hline {$\mathcal{\alpha}$ decay}&$Q_{\alpha}$&$T^{\rm{exp}}_{1/2}$&$Q_{\alpha}^{\rm{WS3+}}$&$T^{\rm{WS3+}}_{1/2}$&${T_{1/2}^{\rm{cal}}}$&$h$&$h^*$&${T_{1/2}^{\rm{pre}}}$&${T_{1/2}^{\rm{gp77}}}$\\ \hline
\endfirsthead
\multicolumn{10}{c}%
{{Table 2. -- continued from previous page}} \\
\hline {$\mathcal{\alpha}$ decay}&$Q_{\alpha}$ &$T^{\rm{exp}}_{1/2}$&$Q_{\alpha}^{\rm{WS3+}}$ &$T^{\rm{WS3+}}_{1/2}$&${T_{1/2}^{\rm{cal}}}$&$h$&$h^*$&${T_{1/2}^{\rm{pre}}}$&${T_{1/2}^{\rm{gp77}}}$\\ \hline
\endhead
\hline \multicolumn{10}{r}{{Continued on next page}} \\
\endfoot
\hline
\endlastfoot
$^{250}$Cm$\to^{246}$Pu&5.17 &$1.45\times10^{12}$& 5.10&  $1.30\times10^{13}$& $6.25\times10^{12}$&0.23&0.33&$2.07\times10^{12}$&$5.10\times10^{13}$\\    
$^{252}$Cf$\to^{248}$Cm&6.22 &$8.61\times10^{7}$& 6.17&  $1.69\times10^{8}$& $1.04\times10^{8}$ &0.83&0.55&$5.73\times10^{7}$&$6.77\times10^{8}$\\    
$^{254}$Cf$\to^{250}$Cm&5.93 &$1.68\times10^{9}$& 5.95&  $2.35\times10^{9}$& $3.25\times10^{9}$ &0.52&0.58&$1.88\times10^{9}$&$2.37\times10^{10}$\\    
$^{256}$Cf$\to^{252}$Cm&5.56 &$1.19\times10^{11}$& 5.59& $2.50\times10^{11}$& $4.13\times10^{11}$&0.29&0.62&$2.54\times10^{11}$&$3.42\times10^{12}$\\    
$^{254}$Fm$\to^{250}$Cf&7.31 &$1.17\times10^{4}$& 7.32&  $7.94\times10^{3}$& $8.93\times10^{3}$ &1.31&0.89&$7.92\times10^{3}$&$4.75\times10^{4}$\\    
$^{256}$Fm$\to^{252}$Cf&7.03 &$1.16\times10^{5}$& 7.05&  $9.58\times10^{4}$& $1.20\times10^{5}$ &0.97&0.90&$1.08\times10^{5}$&$7.03\times10^{5}$\\    
$^{256}$No$\to^{252}$Fm&8.58 &$2.91\times10^{0}$& 8.61&  $1.13\times10^{0}$& $1.41\times10^{0}$ &2.07&1.38&$1.95\times10^{0}$&$5.88\times10^{0}$\\    
$^{258}$No$\to^{254}$Fm&8.15 &$1.20\times10^{2}$& 8.11&  $4.87\times10^{1}$& $3.52\times10^{1}$ &3.41&1.36&$4.77\times10^{1}$&$1.66\times10^{2}$\\    
$^{258}$Rf$\to^{254}$No&9.19 &$1.05\times10^{-1}$& 9.24&  $7.20\times10^{-2}$& $1.03\times10^{-1}$&1.02&2.09&$2.15\times10^{-1}$&$3.87\times10^{-1}$\\    
$^{260}$Rf$\to^{256}$No&8.90 &$1.05\times10^{0}$& 8.92&  $6.23\times10^{-1}$& $6.84\times10^{-1}$ &1.53&1.98&$1.36\times10^{0}$&$2.87\times10^{0}$\\    
$^{260}$Sg$\to^{256}$Rf&9.90 &$1.23\times10^{-2}$& 9.94&  $4.04\times10^{-3}$& $5.12\times10^{-3}$&2.40&3.06&$1.57\times10^{-2}$&$1.78\times10^{-2}$\\    
$^{264}$Hs$\to^{260}$Sg&10.59&$1.08\times10^{-3}$& 10.63& $2.99\times10^{-4}$& $3.67\times10^{-4}$&2.94&3.86&$1.42\times10^{-3}$&$1.19\times10^{-3}$\\    
$^{268}$Hs$\to^{264}$Sg&9.63 &$1.42\times10^{0}$& 9.85&  $2.60\times10^{-2}$& $1.09\times10^{-1}$ &13.03&3.21&$3.49\times10^{-1}$&$4.54\times10^{-1}$\\   
$^{270}$Hs$\to^{266}$Sg&9.07 &$9.00\times10^{0}$& 9.18&  $2.02\times10^{0}$& $4.73\times10^{0}$ &1.90&3.00&$1.42\times10^{1}$&$2.29\times10^{1}$\\    
$^{270}$Ds$\to^{266}$Hs&11.12&$2.05\times10^{-4}$& 10.88& $2.63\times10^{-4}$& $7.39\times10^{-5}$&2.77&3.99&$2.95\times10^{-4}$&$2.32\times10^{-4}$\\    
\noalign{\global\arrayrulewidth1pt}\noalign{\global\arrayrulewidth0.4pt} \multicolumn{10}{c}{\rm{$^{294}$Og $\to{}^{290}$Lv $\to{}^{286}$Fl $\to{}^{282}$Cn}}\\
$^{294}$Og$\to^{290}$Lv&11.84&$1.15\times10^{-3}$&11.97&$1.09\times10^{-4}$& $2.24\times10^{-4}$&5.14&1.86&$4.16\times10^{-4}$&$8.68\times10^{-4}$\\  
$^{290}$Lv$\to^{286}$Fl&11.01&$8.00\times10^{-3}$&10.88&$1.15\times10^{-2}$& $5.52\times10^{-3}$&1.45&2.36&$1.30\times10^{-2}$&$2.40\times10^{-2}$\\  
$^{286}$Fl$\to^{282}$Cn&10.37&$3.50\times10^{-1}$&9.94& $9.13\times10^{-1}$& $6.14\times10^{-2}$&5.70&2.77&$1.70\times10^{-1}$&$2.88\times10^{-1}$\\  
\noalign{\global\arrayrulewidth1pt}\noalign{\global\arrayrulewidth0.4pt} \multicolumn{10}{c}{\rm{$^{296}$Og $\to{}^{292}$Lv $\to{}^{288}$Fl $\to{}^{284}$Cn}}\\
$^{296}$Og$\to^{292}$Lv&{}&{}&         11.62&$6.39\times10^{-4}$&  $6.39\times10^{-4}$&{}&1.71&$1.09\times10^{-3}$&$2.64\times10^{-3}$\\     
$^{292}$Lv$\to^{288}$Fl&10.78&$2.40\times10^{-2}$&10.92&$8.36\times10^{-3}$&  $1.93\times10^{-2}$&1.25&2.21&$4.25\times10^{-2}$&$9.04\times10^{-2}$\\    
$^{288}$Fl$\to^{284}$Cn&10.07&$7.50\times10^{-1}$&9.47& $2.14\times10^{1}$&  $3.77\times10^{-1}$&1.99&2.63&$9.90\times10^{-1}$&$1.94\times10^{0}$\\

\end{longtable}
\end{center}

\begin{multicols}{2}

\begin{center}
\includegraphics[width=8.5cm]{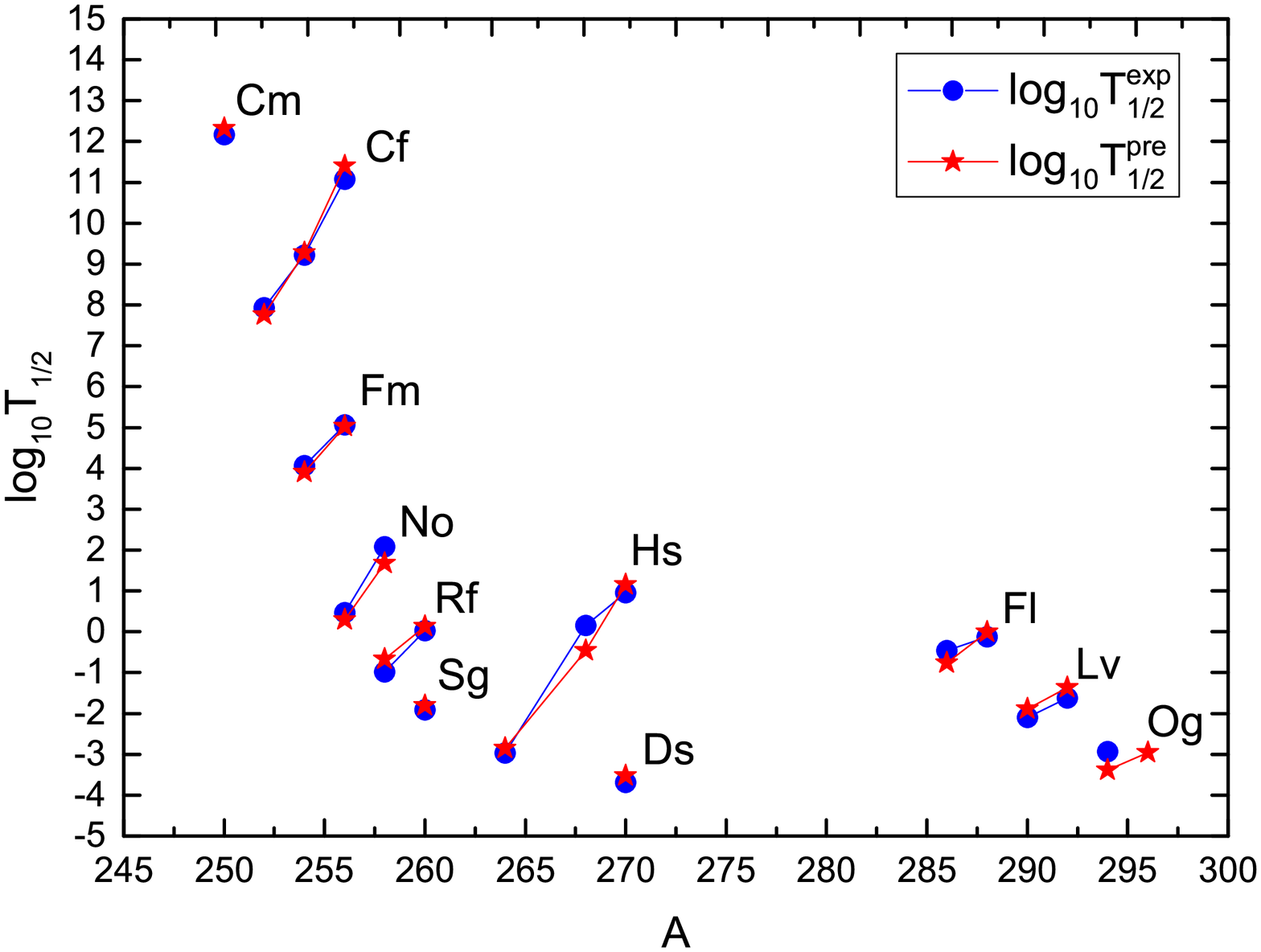}
\figcaption{\label{figure1}(color online) Logarithmic half-lives of experimental data and predicted ones. The blue circle and red star denote the experimental half-lives $T_{1/2}^{\rm{exp}}$, and predicted results $T_{1/2}^{\rm{pre}}$, respectively.}
\end{center}

\begin{center}
\includegraphics[width=8.5cm]{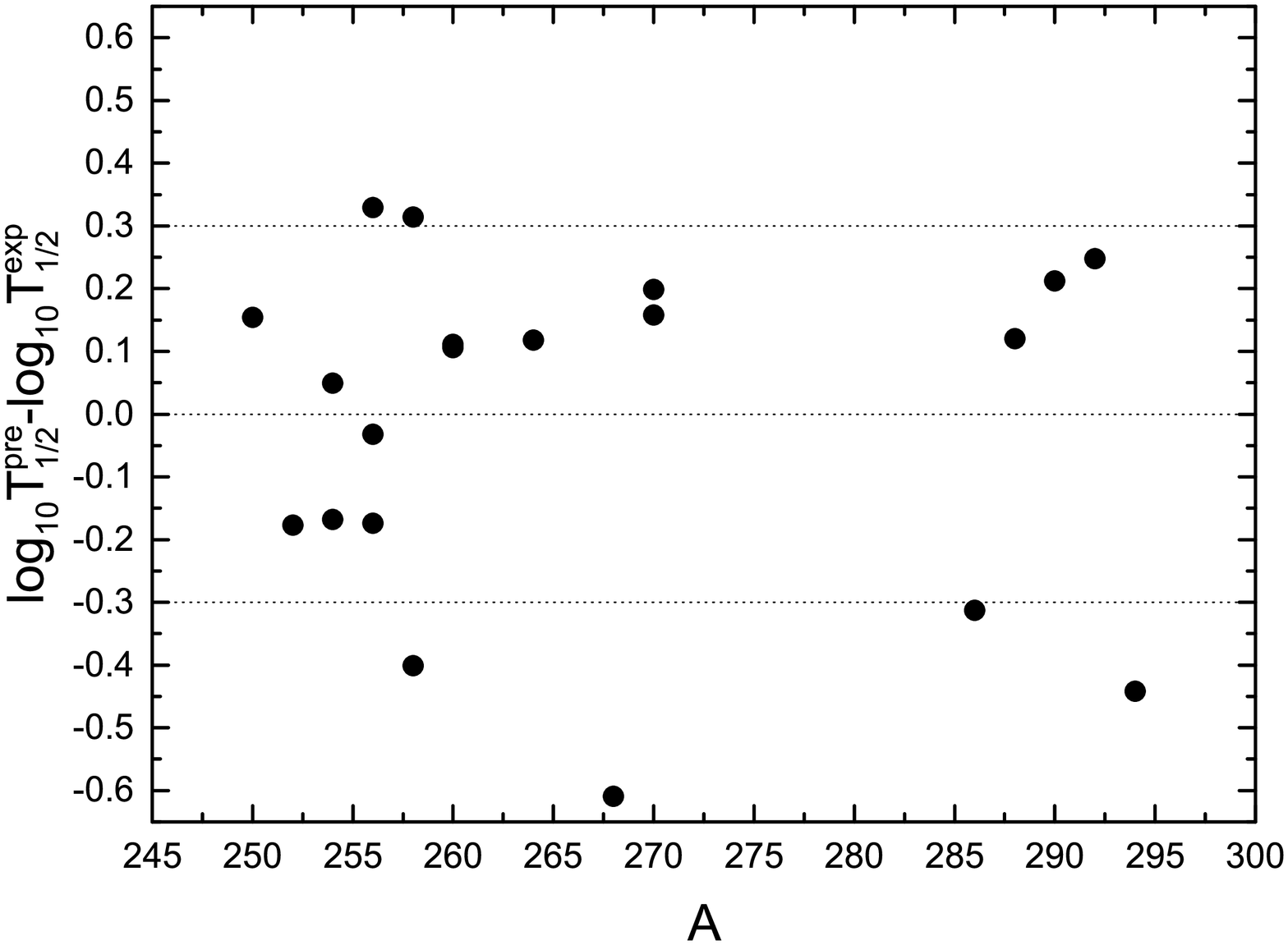}
\figcaption{\label{figure2} The logarithmic differences between $T_{1/2}^{\rm{pre}}$ and $T_{1/2}^{\rm{exp}}$.}
\end{center}

\section{Summary}

In summary, we predict $\mathcal{\alpha}$ decay half-life of $^{296}$Og and systematically calculate $\mathcal{\alpha}$ decay half-lives of all 20 even-even nuclei in the same region with nucleus $^{296}$Og from $^{250}$Cm to $^{294}$Og by adopting TPA, and extract corresponding $\mathcal{\alpha}$ decay hindrance factors as well as a new set of parameters of hindrance factors considering the shell effect. Our calculations i.e. $T_{1/2}^{\text{pre}}$ considering the hindrance factor correction can well reproduce the experimental data. The predicated $T_{1/2}$ of $^{296}$Og is 1.09 ms within a factor of 5.12. This work will be used as a reference for synthesizing nucleus $^{296}$Og.
\end{multicols}
\vspace{-1mm}
\centerline{\rule{80mm}{0.1pt}}
\vspace{2mm}

\begin{multicols}{2}

\end{multicols}

\clearpage
\end{CJK*}
\end{document}